# A NEW EMAIL RETRIEVAL RANKING APPROACH

Samir AbdelRahman[1], Basma Hassan[2] and Reem Bahgat[1]
[1]Department of Computer Science, Faculty of Computers and Information, Cairo University, Giza, Egypt
s.abdelrahman@fci-cu.edu.eg , r.bahgat@fci-cu.edu.eg
[2]Department of Computer Science, Faculty of Computers and Information, Fayoum University, Fayoum, Egypt
bhk00@fayoum.edu.eg

### ABSTRACT

*Email Retrieval task has recently taken much attention to help the user retrieve the email(s) related to the submitted query. Up to our knowledge, existing email retrieval ranking approaches sort the retrieved emails based on some heuristic rules, which are either search clues or some predefined user criteria rooted in email fields. Unfortunately, the user usually does not know the effective rule that acquires best ranking related to his query. This paper presents a new email retrieval ranking approach to tackle this problem. It ranks the retrieved emails based on a scoring function that depends on crucial email fields, namely subject, content, and sender. The paper also proposes an architecture to allow every user in a network/group of users to be able, if permissible, to know the most important network senders who are interested in his submitted query words. The experimental evaluation on Enron corpus prove that our approach outperforms known email retrieval ranking approaches.*

## 1. INTRODUCTION

Information Retrieval (IR) [1] is an interdisciplinary domain that searches for documents, information, or metadata in a specific data storage as the IR techniques proved to enhance domains like Search Engines, Question Answering, Information Extraction, and Recommenders. The IR system effectiveness depends on two factors: (1) the quality of retrieval, i.e. the number of retrieved documents that are relevant to the submitted query; and (2) the quality of the ranking model, i.e. to what extent the order of these documents conformed with their actual importance to the query.

Email Retrieval is one of the most important IR branches in which the data storage is the user e-mailbox that includes set of emails. An email is a semi-structured document composed of many fields, such that each field describes a piece of information. They could be mainly categorized into five types: (1) The Date field to hold the email date; (2) The Content field to contain the email content; (3) The fields having the representative words, namely Keywords and Subject fields; (4) The fields of email actors that hold the email sender, i.e. From, or email recipients, i.e. To, Cc, and Bcc; and (5) The email handlers, namely Importance, Priority, and Security, to hold information about how the mail server and the recipients may deal with the corresponding email.

All above email fields are exploited to extract hidden useful information from the user e-mailbox. For example, Xiang [2] presented a system that automatically clusters the emails in the e-mailbox according to their content and subject fields. Cselle et.al [3] used clustering techniques on subject, content, and actor email fields to group emails by topics. Erera and Carmel [4] also used the same email fields with classification techniques for detecting conversations in the user e-mailbox. Castro and Lopes [5] presented a visualization system for

email retrieval based on the email content field using visualization techniques. Yee et.al [6] used the actor email fields to extract social network data from email logs.

All the above mentioned work does not use email fields to boost Email Retrieval Ranking. The existing Email Retrieval systems rank the emails either by using some user clues or some predefined criteria, which sort the retrieved emails based on the email fields like Date, Subject, From, and Priority. Unfortunately, the email may not include the user clue, though its theme is related to such a clue, and vice versa. Moreover, using predefined user criteria for email ranking do not help the user to know which emails are crucial to the submitted query words, because the ranking process becomes query independent.

Among all email fields, people consider Subject, Content, and From are the most expressive fields to present the most valuable email information. From the user perspective, the email subject and content words, which determine the email theme, are very important. Moreover, the email sender, whose emails are mostly related to the submitted query words, should be considered as the email theme specialist. Based on these user perspectives, we propose a simple scoring function as a combination of email subject, content, and sender. This function is to weigh the retrieved emails relevant to the submitted query such that each retrieved email should include the whole query words. The retrieved emails are then sorted in descending order by their weights. The proposed scoring function calculations are done to gain accurate weights, which mostly express the real importance of the corresponding emails to the submitted query. To do that, we basically utilize the email subject, content, and sender as follows:

**Subject:** Email subject presents the representative email words (the email theme). We increase the weight of the email having subject that includes all query words.

**Content:** Email content typically describes the email objective. We divide the email content into a set of segments, namely the unquoted part and the quoted parts. In addition, we make current email threading scheme be able to handle such segments effectively. The email content weight is calculated based on two factors. First, the query word importance, where the importance is increased proportionally to the number of times that the word appears in the email offset by the word email frequency in the e-mailbox. Second, the segment includes this word such that the unquoted segment is the most effective email part because it states some recent debates regarding the quoted parts.

**Sender:** Email sender, from the user point of view, is the foremost email actor interested in the email theme. The user also considers the sender to be a theme specialist if the latter emails are mostly related with that theme. We weigh the email sender from this viewpoint. Moreover, the sender weight is boosted if he is one of the query theme specialists in the network. To do that, the proposed email network architecture allows the network server to have all user names coupled with their most e-mailbox frequent words. By exchanging such knowledge among all network users, if permissible, the sender weight calculation becomes more accurate and expressive.

To accomplish our goals, the following two decisions were made:

1) Enron Corpus [7] was chosen as a benchmark data for our experiments. Indeed, up to our knowledge, there is no Email Ranking test collection available. Therefore, we set up our own test collection manually.
2) Up to our knowledge, there is no Email Retrieval Ranking baseline system available to compare with its performance. Therefore, we implemented some current Email Retrieval Ranking approaches and we used some ready-made interfaces of known commercial email systems. We tested our approach heuristics against their results.

In comparison with the current Email Retrieval Ranking approaches, the proposed approach contributions are as follows:

1) Its proposed heuristics are simple. They are based on realistic human thinking when they review their emails, while all current ranking approaches hypothetically assume that people know well the best ranking criterion to acquire emails that correctly match their query.
2) It utilizes the hierarchical and referential structures of emails effectively in ranking the retrieved emails. This is acquired through our proposed email threading algorithm. Unfortunately, some of the current ranking approaches use email threads for only two purposes: 1) retrieving the whole thread if any of its emails is relevant to the query; therefore, some irrelevant emails may be retrieved as well; 2) applying thread based ranking for the retrieved emails such that the retrieved emails are grouped in threads.
3) It considers the expertise of the shared network email users regarding the query theme in the ranking function calculations. In addition, it introduces such knowledge to the user as system recommendations. To date, up to our knowledge, all the current ranking approaches constrain the search scope to the user e-mailbox only. Therefore, none of these approaches benefit from the network users expertise in email retrieval ranking.

The remainder of this paper is organized as follows: Section 2 describes some current email ranking and threading approaches. Section 3 states the problem definition and notations. Section 4 explains the email threading algorithm used in the proposed approach. Section 5 submits the proposed approach architecture and design. Section 6 states the experimental analysis and results to show the proposed approach effectiveness. Section 7 concludes the work and focuses on some important future directions.

## 2. RELATED WORK

### 2.1. Email Retrieval Ranking

Perkiö et.al [8] proposed an email retrieval ranking approach that utilizes a combination of four facets, namely the search attribute, the topic, the sociality, and time-series analysis. The approach is totally based on clues that the user submits with the query keywords. For example, if the user wants to search his e-mailbox for emails that include "George Bush" words and talk about politics, then he may submit "politics" as a clue of "George Bush" words. The system first segments each email to a number of blocks. Second, all emails that include the keywords, i.e. "George Bush", are retrieved. Third, for each email, the system calculates the ratio of the words that appear with "politics" in the same blocks to the repetitions of these words in the emails. Finally, the email is scored by this ratio and the retrieved emails are sorted in descending order accordingly. However, this approach may lose some emails that talk about politics and they do not include the word "politics" at the same time. Moreover, some emails may be highly ranked; nevertheless, they have many "politics" word occurrences to debate subjects that are not related to political themes.

López et.al [9] represent e-mailboxes as the World Wide Web and interpret e-mailbox searching as the search engine problem. The user e-mailbox is transferred to a website. Each website consists of the user home page (the user profile), a page for each sent email, and a page to include the links of all received emails. Google Page Ranking algorithm is applied to rank the users by their citations to each other. Moreover, an email is considered important if it is sent to important people. Unfortunately, this email ranking is query independent; hence, the email orders are not related to the submitted query words.

Weerkamp et.al [10] presented a study to use messages through public discussion lists. Its aim is to expand the submitted query by the most likelihood clues (contextual information) related to this query, which are deduced from the retrieved emails of previous user similar queries. This approach aims to retrieve and sort the discussion list messages using these clues coupled with some prior query independent features for each discussion message details, such as its length, its quality (linguistic error free), or its thread length. This approach may work properly only on discussion forums, as forum people always tend, in their messages, to repeat the clues related to the discussion theme. Moreover, the message prior information features, which are used to calculate the clue probability, may not be good indicators for email clue extraction. That is because most of the emails have linguistic errors and the length of the user emails or threads does not express the significance of their corresponding clues.

Gmail system uses the email date as the default ranking attribute. It may use one or more other email fields, such as email actors and subject, to impose some email constraints on the user e-mailbox to filter the retrieved emails. Windows Email system uses each mentioned field, which is selected by the user, solely to rank the retrieved emails.

### 2.2. Email Threading

Yeh and Harnly [11] proposed an email threading technique to cluster emails within time interval limits based on their subject similarities. Therefore, the resultant thread consists of the emails occurred within a specific period having similar subject. Each thread is represented by a tree, in which the email is a parent and the reply or forward email to this parent is considered as its child. This technique needs to re-cluster the emails if a new email arrives to the user e-mailbox. Therefore, we adapted it to practically support our ranking approach (Section 4).

To date, up to our knowledge, all commercial email systems, such as Gmail or Windows Email, use the email subject field to cluster the emails into threads. This approach considers each thread as a compound of related emails having similar subject. Therefore, the whole thread, which includes the query words, is always retrieved even if some of its emails are not actually related to these words; consequently, the overall system precision may be decreased.

## 3. PROBLEM DEFINITION

This paper is interested in ranking the retrieved emails ($E_R$), from the user e-mailbox ($M$), that are relevant to the submitted query ($Q$). The paper aims to propose a ranking solution that utilizes some heuristics to sort the retrieved emails that are close to people's thinking (Section 5). To do that, the e-mailbox is considered as a set of many emails, such that each email ($e$) is composed of many email documents ($ed$), i.e. main body ($ed_b$) and set of quotations segments ($ED_{QT}$) (Section 4). The emails having similar subject are grouped into a single email thread ($T$). To carry out our approach, the lexemes, set of words ($w$) and subject ($sb$), are recognized; therefore, we have thread subject words ($T_{sb}$), email subject words ($e_{sb}$), and query words ($Q_w$). Finally, our retrieval algorithm uses the submitted query to get $E_R$, $E_R \subseteq M$, such that we have retrieved threads ($T_R$) and retrieved email documents ($ED_R$) from whole email documents ($ED_M$) in the e-mailbox.

## 4. THE PROPOSED EMAIL THREADING ALGORITHM

Email Thread is a sequence of emails having identical subject field and they are responses to the initial email. The email may be single document or a composition of many documents, presented by the main body text (the unquoted part) and the quotations. While each quotation is a separate document, the main body is the actual text written by the email sender.

Most of the current email systems consider the whole email body as a single unit document. In this case, some irrelevant emails are retrieved, especially if some query terms are scattered among the main body and the quotations. For example, Figure 1 depicts a sample Enron corpus email retrieved for the query "Conference Call" by the current email systems, such as Gmail and Windows Mail. The email appears in the top of the retrieved email list; though, it is irrelevant to such a query. To overcome this problem, we chose the main body to be the unit document for indexing and we used inverted index [1]. Moreover, the document representation model presented by Broder et al. [12] was adapted and used to present the quotations in our inverted index.

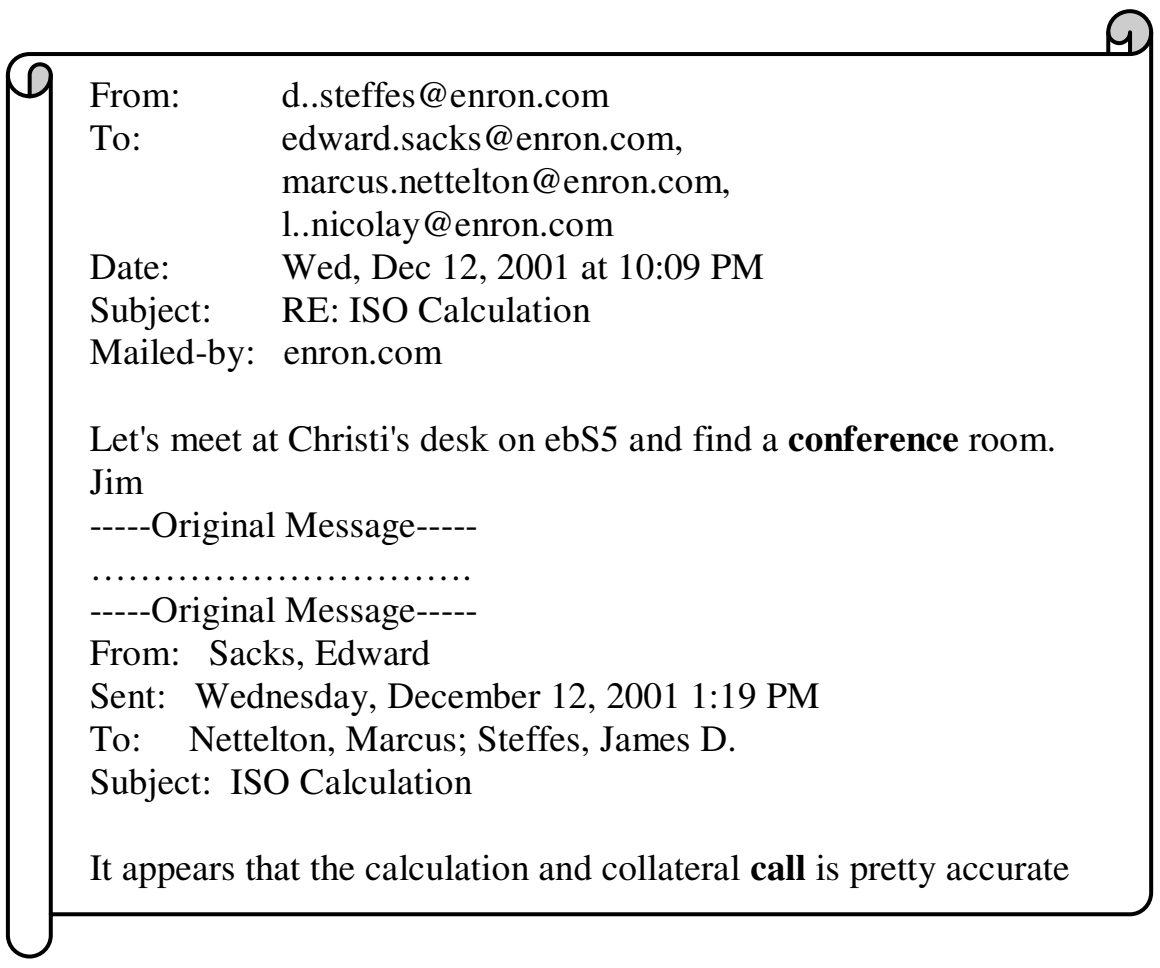
From: d..steffes@enron.com
To: edward.sacks@enron.com, marcus.nettelton@enron.com, l..nicolay@enron.com
Date: Wed, Dec 12, 2001 at 10:09 PM
Subject: RE: ISO Calculation
Mailed-by: enron.com

Let's meet at Christi's desk on ebS5 and find a **conference** room.
Jim
-----Original Message-----
…………………………
-----Original Message-----
From: Sacks, Edward
Sent: Wednesday, December 12, 2001 1:19 PM
To: Nettelton, Marcus; Steffes, James D.
Subject: ISO Calculation

It appears that the calculation and collateral **call** is pretty accurate

Figure 1. Sample Email from Enron Corpus

In our model, the tree is constructed using the email parent-child relationship, in which each child email includes the main body of its parent as a quotation. Hence, if the parent is relevant to the submitted query then its children are relevant too.

For the Email Retrieval, when a query is submitted, a set of emails are retrieved from the user e-mailbox. Each email retrieved contains the submitted query words in (1) the main body, (2) the quotation parts, or (3) the subject field. Since only the main body is indexed, as been previously mentioned, the index is used in retrieving the first type only, while the others are retrieved using email threads.

Our email threading algorithm aims to construct the email threads of the user e-mailbox using trees. To achieve this, the email thread re-assembly algorithm presented by Yeh and Harnly [11] was adapted. In our algorithm, the node of the thread tree was chosen to be the email document instead of the email itself.

In the e-mailbox, there may exist an email whose some of its original emails of related quotations do not exist in the e-mailbox. Thus, when receiving, sending, or saving a new email, if any of its documents does not match with a node document in the email threads, this document is added to the thread that it belongs to. Then, it is indexed.

The main functions of the email threading algorithm are addition, deletion, update, and retrieval. Since the most repetitive operations done on the user e-mailbox are addition and retrieval, we focus only on them in the following explanation:

***Function1: The Email Addition***

When a new email ($e$) is received, sent, or saved, this function inserts $e$ in the thread that $e$ belongs to, if found, or it creates a new thread having $e$ only. Then, the email is preprocessed in two steps, namely *Subject Normalization* and *Body Segmentation*. Subject normalization is the task of removing any prefixes in the email subject like, “Re:”, “Fw:”, and “Fwd:” to obtain a normalized subject ($e_{sb}$) for this email. Body segmentation is the task of segmenting the email body into main body ($ed_b$) and set of quotations ($ED_{QT}$). Each quotation ($ed_i$) is extracted away from the quotation patterns and headers. Figure 2 illustrates the preprocessing done on a sample

email that consists of three documents, the $ed_b$, i.e. $ed_0$, and the two quotations, i.e. $ED_{QT} = \{ed_1, ed_2\}$. The email $e_{sb}$ is '*Revised Daily Notice*'.

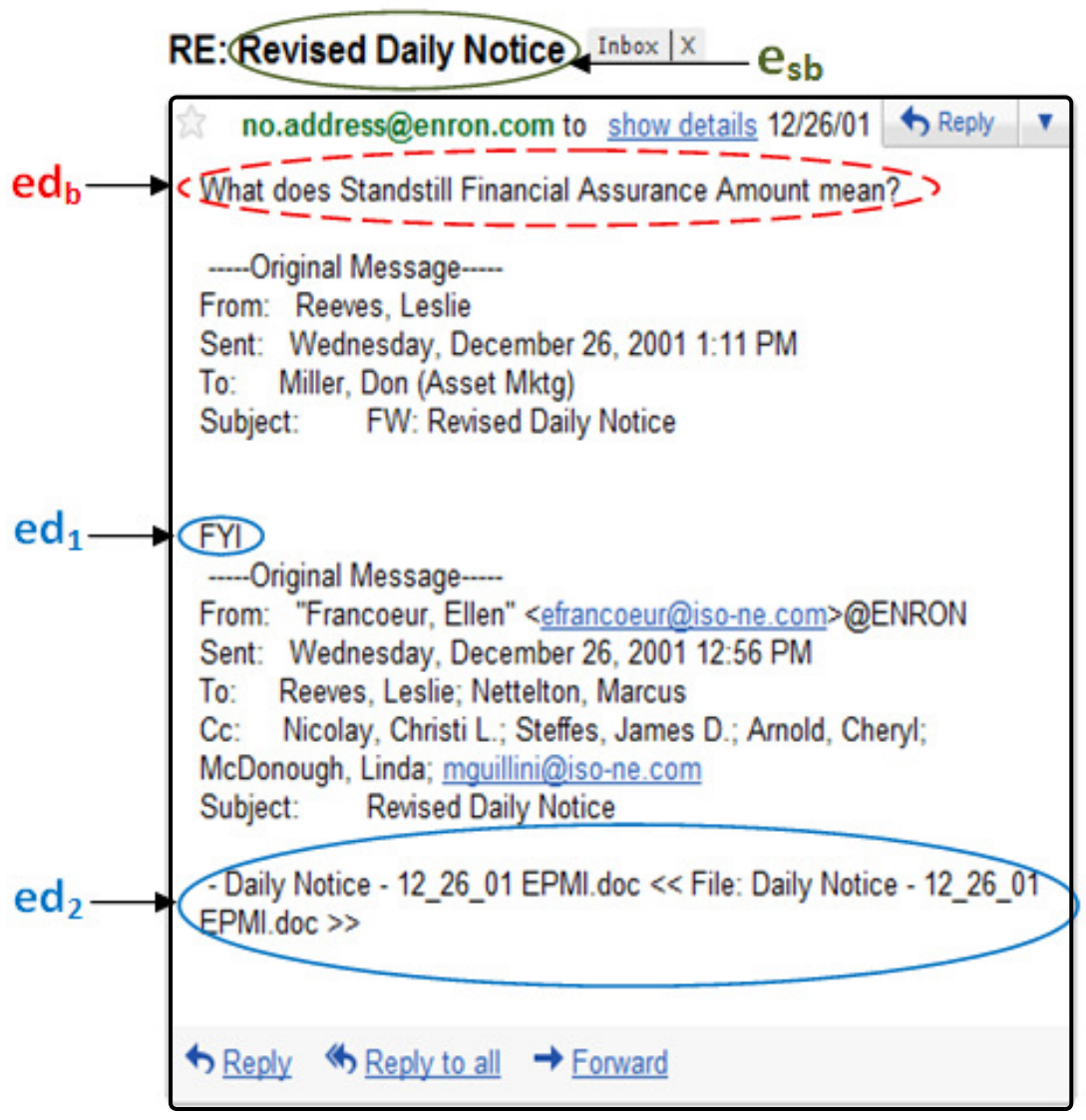

Figure 2. Email Preprocessing Example

After preprocessing, the function steps are as follows:

**Step1**,

All threads whose subject is similar to $e_{sb}$ are retrieved ($T_S$).

If no thread is found with subject $e_{sb}$, i.e. $T_S = \varnothing$, a new thread $T_n$ is created with $T_{sb} = e_{sb}$, and every $ed_j$, $ed_j \in e$, is inserted in $T_n$, in which, the initial document, i.e. the oldest quotation $ed$, is inserted. Then, the reply document to it is inserted as its child node, and so on until inserting the main body $ed_b$. Therefore, $ed_b$ becomes the leaf of $T_n$ and the initial document becomes the tree root.

Else, go to Step 2.

**Step2**,

For each thread $T_j$, $T_j \in T_S$, the path $P$ is retrieved where $P$ is the longest path of $T_j$ nodes, starting by the node containing $ed$ with content similar to the oldest quotation in $e$. The ending node in $P$ is the node, in a lower tree level, containing $ed$ with the content similar to the more recent, higher level, $ed$ in $e$.

Based on the length of both $P$ and $e$, denoted by $n_p$ and $n_e$ respectively, one of the following four cases is satisfied.

***Case1:*** *"All Quotations Matched"*

The number of nodes in $P$ equals to the number of quotations in $e$ ; i.e. $n_p = n_e - 1$.

In this case, each $ed$ in $ED_{QT}$ exists in $T_j$ ; i.e. $e$ is a reply to an existing email. Therefore, $ed_b$ is inserted in $T_j$ as a child to the end node in $P$.

***Case2:*** *"Some Quotations Matched"*

The number of nodes in $P$ is less than the number of quotations in $e$ ; i.e. $n_p < n_e - 1$.

In this case, only some of *ed*s in $ED_{QT}$ are found in $T_j$ , i.e. $P \subseteq ED_{QT}$. This means that *e* is a forward to an email that one or some of its eldest replies exist in the user e-mailbox, but the newest replies are not found. Therefore, for every $ed_i$ , $ed_i \in ED_{QT}$ and $ed_i \notin P$ , $ed_i$ is inserted in $T_j$ . Finally, $ed_b$ is inserted in $T_j$ as well.

***Case3:*** "*All Email Documents Matched*"

The number of nodes in *P* equals to the number of documents in *e*, including $ed_b$; i.e. $n_p = n_e$ .

In this case, each *ed* in *e* exists in $T_j$, including $ed_b$. This means that *e* is a delayed email, i.e. it is received lately, and there exists a subsequent reply or a forward email to it in the user e-mailbox. Therefore, no new *ed* is needed to insert.

***Case4:*** "*No Document Matched*"

$n_p = 0$ .

In this case, no path found matches with *ed*s of *e* in $T_j$. Therefore, a new thread is created for *e* with $T_{sb} = e_{sb}$ ; similar to Step1 when no thread found with $T_{sb} = e_{sb}$ .

**Example:**

Figures (3 - 6), illustrate some email threading operations to explain the four possible cases in Step 2. The email in Figure 2 is received and preprocessed. Then after applying Step1, $T_j$ is one of the threads retrieved, i.e. $T_j \in T_S$ ; $T_{sb}$ = '*Revised Daily Notice*'. For each case in the figures, the matched path *P* in $T_j$ is circled and the action done is highlighted in red.

Case 1 is the most common behavior that occurs in the user e-mailbox, in which the email is a direct reply to the original email of the first quotation; i.e. $ed_1$. In this case, the two quotations of the email are matched with nodes in *P*, so $ed_b$ is inserted as a child node to $ed_1$. Figure 3 illustrates this case.

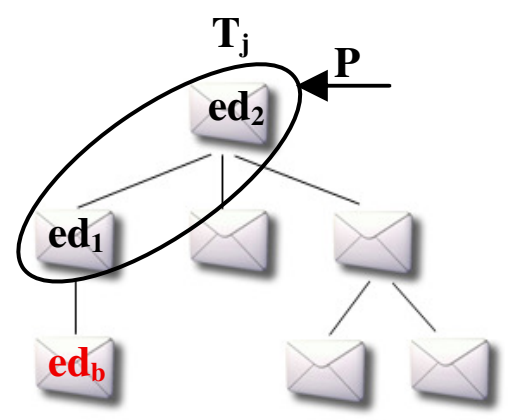

Figure 3. Case 1: "All Quotations Matched"

Case 2 considers the situation where the email sender forwards it to the user after some series of replies, or the user was added as a recipient when replying to another email. In both situations, only the initial email of the thread exists in the user e-mailbox, which is $ed_2$. Therefore, all non-matched quotations, which exist in the email, are inserted in $T_j$, each one is considered as a child to the document that it replies, so $ed_1$ is inserted as a child to $ed_2$, and $ed_b$ is inserted as a child to $ed_1$. Figure 4 illustrates this case.

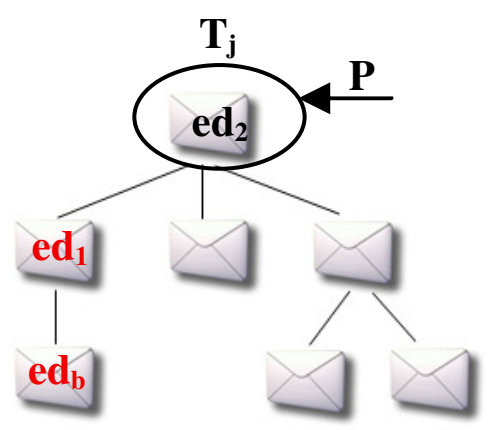

Figure 4. Case 2: "Some Quotations Matched"

Case 3 rarely happens in the user e-mailbox, where the email is received late after its reply has arrived. In this situation, all documents of the email should be inserted in the thread upon receiving the reply. Therefore, this email adds nothing to the thread. Figure 5 illustrates this case.

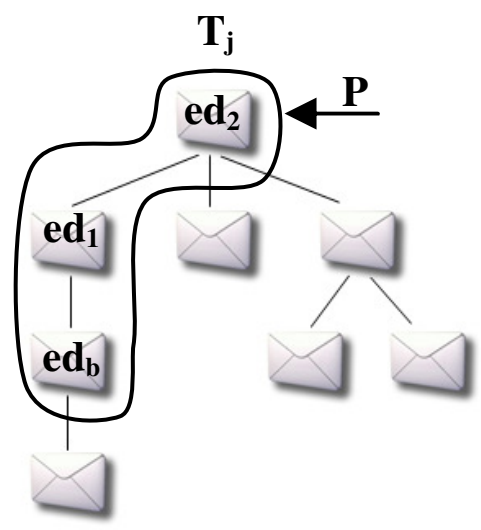

Figure 5. Case 3: "All Email Documents Matched"

The last case, Case 4, considers the situation when the email has subject similar to another email in the e-mailbox but it is not related to it, i.e. the emails share only one subject but do not share the body content. In this case, the email is an initial email for a new thread $T_n$ with $T_{sb}$ = '*Revised Daily Notice*'. Figure 6 illustrates this case.

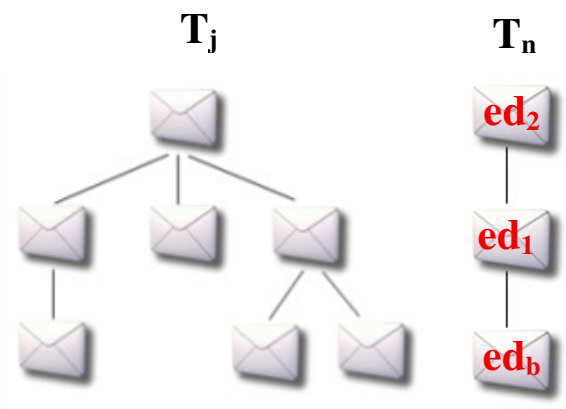

Figure 6. Case 4: "No Document Matched"

***Function2: The Email Retrieval***

On the query submission, two inputs are given to this procedure: (1) query words $Q_w$, and (2) set of email documents that are retrieved by index ($ED_I$). Then it returns:

- All threads $T_R$ that contain $Q_w$ in their subject, i.e. $Q_w \subseteq T_{sb}$ for each $T$ in $T_R$ .
- $ED_{child}$ set for each $ed_j$ in $ED_I$ , where $ED_{child}$ are descendants of $ed_j$ in the thread tree.

## 5. The Proposed Ranking Approach

As been previously mentioned, the proposed ranking approach uses a scoring function that consists of three of the most expressive email fields, i.e. subject, content, and sender. The proposed approach considers the retrieved email to be more relevant to the submitted query if the email content, including the email subject, contains the query theme, also, if the email comes from a sender whose most received emails are interested in the query theme. Based on such heuristics, the emails are scored and ranked in descending order. In the following subsections, the proposed system architecture coupled with the score function principles and heuristics are described.

### 5.1. The Subject Scoring

Email subject is the email title that usually presents the email theme abstract. Heuristically, the user considers the subject words to be the main email clues. Therefore, an email is considered highly relevant to the query if the query words are included in the email subject. Moreover, since the email thread consists of all emails having similar subject, all threads whose subject matches the query words are retrieved ($T_R$) (Section 4). Therefore, all the emails included in the

threads in $T_R$ are added to $E_R$ . Finally, Equation (1) is used to calculate subject score *TScore* for each email retrieved.

$$TScore(e) = \begin{cases} 1, & \text{if } e \in T_R \\ 0, & \text{Otherwise} \end{cases} \quad ,\forall\ e \in E_R \tag{1}$$

## 5.2. The Content Scoring

Email content is the email theme. The user considers email terms to be important words if they are frequently used in his e-mailbox. Moreover, people use *TF-IDF* [1] as a ranking factor, in which the term that is frequently used in fewer documents should get higher weight than the others. Therefore, Equation (2) adapts this factor as good heuristics to weigh the document content words.

$$TF - IDF(tr_i, ed_j) = TF_{i,j} \times \log \frac{N}{DF_i} \tag{2}$$

$tr_i$ is the i[th] term (word) $\in j^{th}$ *ed* , where $ed_j \in ED_M$

$TF_{i,j}$ is the frequency of $tr_i$ in $ed_j$

$DF_i$ is the number of any $ed_j \in ED_M \ni tr_i \in ed_j$

N = | $ED_M$ | ; i.e. number of $ed \in M$

We use the vector space model [1] by assigning vectors for the submitted user query and all the documents retrieved, such that each vector component is the term *TF-IDF*. Then, we use Cosine Similarity [1] between the query and each document to measure the latter weight (Equation (3)).

$$Sim(ed_j, Q) = \frac{ed_j \bullet Q}{| ed_j | \bullet | Q |} \quad ,\forall\ ed_j \in ED_R \tag{3}$$

We aim to score the retrieved email, and since it is composed of many documents, their similarities to the query (Equation (3)) are summed. Additionally, the location of the email document in the email presents its importance to the query from the user perspective; i.e. the main body ($ed_0$) is the most important, while the eldest quotation ($ed_{n-1}$) is the least important. Therefore, the email content score *CScore* (Equation (4)) is calculated as the weighted documents similarities to the submitted query based on the document level within the email. Heuristically, we consider that the document similarity should be decreased by half its similarity if located in the higher level.

$$CScore(e) = \sum_{j=0}^{n-1} 0.5^j \times Sim(ed_j, Q) \quad ,\forall\ e \in E_R \tag{4}$$

Where n = | e | ; i.e. number of $ed \in e$ .

## 5.3. The Sender Scoring

Email sender (*s*) is the most email actor interested in the email theme. Usually, the user considers the email sender to be interested in the email theme if most of his emails are related with such a theme. Moreover, the sender is believed to be highly professional in some user themes if the sender's most frequent emails are related to the theme while the others rarely use them. According to such heuristics, *TF-IDF* (Equation (2)) is adapted to score the email senders (*S*). To do that, each sender emails are considered a single email. Therefore, the term that is

frequently used in fewer emails (senders) is highly weighted, and vice versa. Then, we use the vector space model to present each sender and query as vectors, such that each vector component is the term *TF-IDF*. Finally, Equation (6) is applied to weigh the email using its sender score *SScore*.

$$Sim(s,Q) = \frac{s \bullet Q}{|s| \bullet |Q|} \qquad ,\forall\, s \in S \qquad (5)$$

$$SScore(e) = Sim(s,Q) \qquad ,\forall\, e \in E_R \qquad (6)$$

### 5.4. The Proposed Scoring Function

The user always searches the email whose content is more relevant to the submitted query. Moreover, he prefers to retrieve first the email that was sent by the query theme specialist. Therefore, the email subject score (Equation (1)) is added to the content score (Equation (4)), and this summation is weighted by the email sender score (Equation (6)) to apply the mentioned user preference (Equation (7)).

$$Score(e) = SScore(e) \times (CScore(e) + TScore(e)) \qquad ,\forall\, e \in E_R \qquad (7)$$

### 5.5. The Proposed Architecture

People socially try to get benefit from their network shared data. Therefore, we propose an architecture that exploits the networking communication nature among email users to acquire information about each user interests, if permissible. Figure 7 presents an abstract view of the proposed architecture. It shows that the architecture consists of one server and a set of clients (user e-mailbox), such that the server coordinates the exchangeable information among the email users. The server, also, stores and updates all network users e-mailbox words coupled with *TF-IDF* for each word.

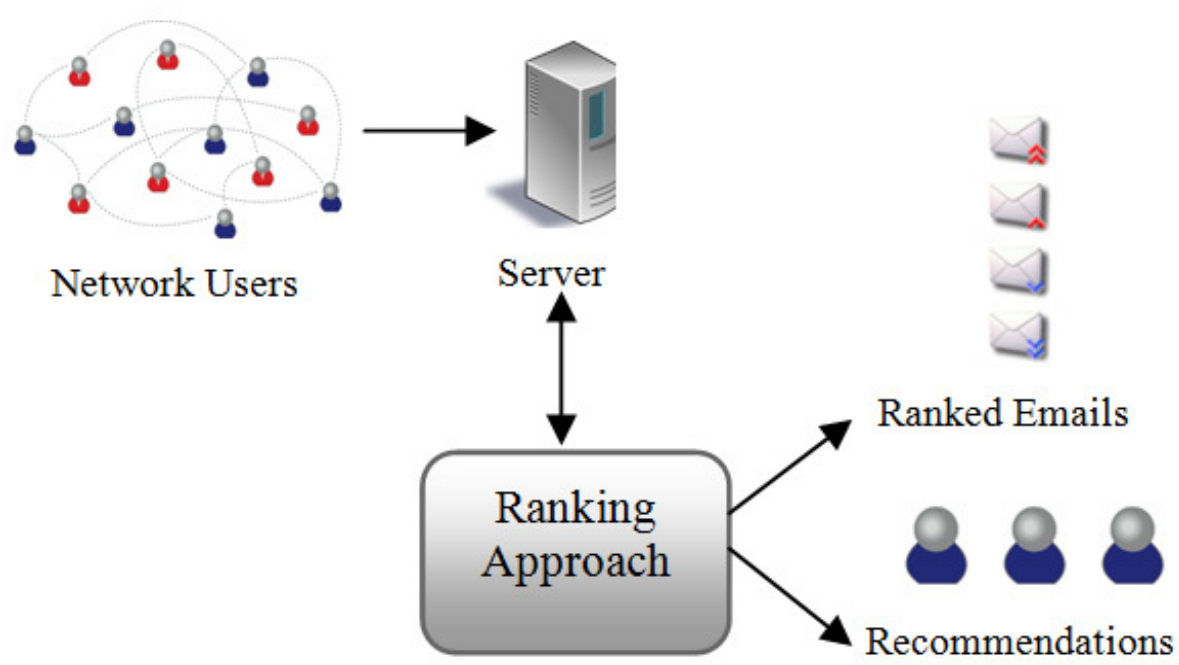

Figure 7. The Proposed Networking Architecture

Each client (Figure 8) is mainly composed of four autonomous objects: (1) *Sender-O* as an interface object between the server and Search-O, (2) *Search-O* as a link among all other client objects to get user ranked emails and recommendations, (3) *Index-O* as an interface object of the user e-mailbox inverted index, and (4) *Thread-O* as an object to handle email threading algorithm functions (Section 4). A typical client-server communication scenario may be highlighted as follows:

1) The user submits a query. Then, Search-O tokenizes the query words and sends them to Sender-O.
2) Sender-O communicates with the server to get all senders interested in the query words. The object retrieves a list of senders names and their *TF-IDF* of each query word.

3) While Sender-O doing his task, Search-O sends the query words to Thread-O and Index-O, to get the related information from them.
4) Thread-O applies email threading retrieval function (Section 4) to obtain all threads that include the query words.
5) Index-O retrieves all documents that contain the query words, from the e-mailbox index.
6) Search-O receives all other client objects (Steps 2, 4, and 5). It recommends all senders retrieved from Sender-O and did not appear in the user contact list; also, it uses the others to rank the retrieved emails using Equation (7).
7) Search-O sends the updates of the user *TF-IDF* factors to Sender-O.
8) Sender-O communicates with the server for new updates (Step 7).

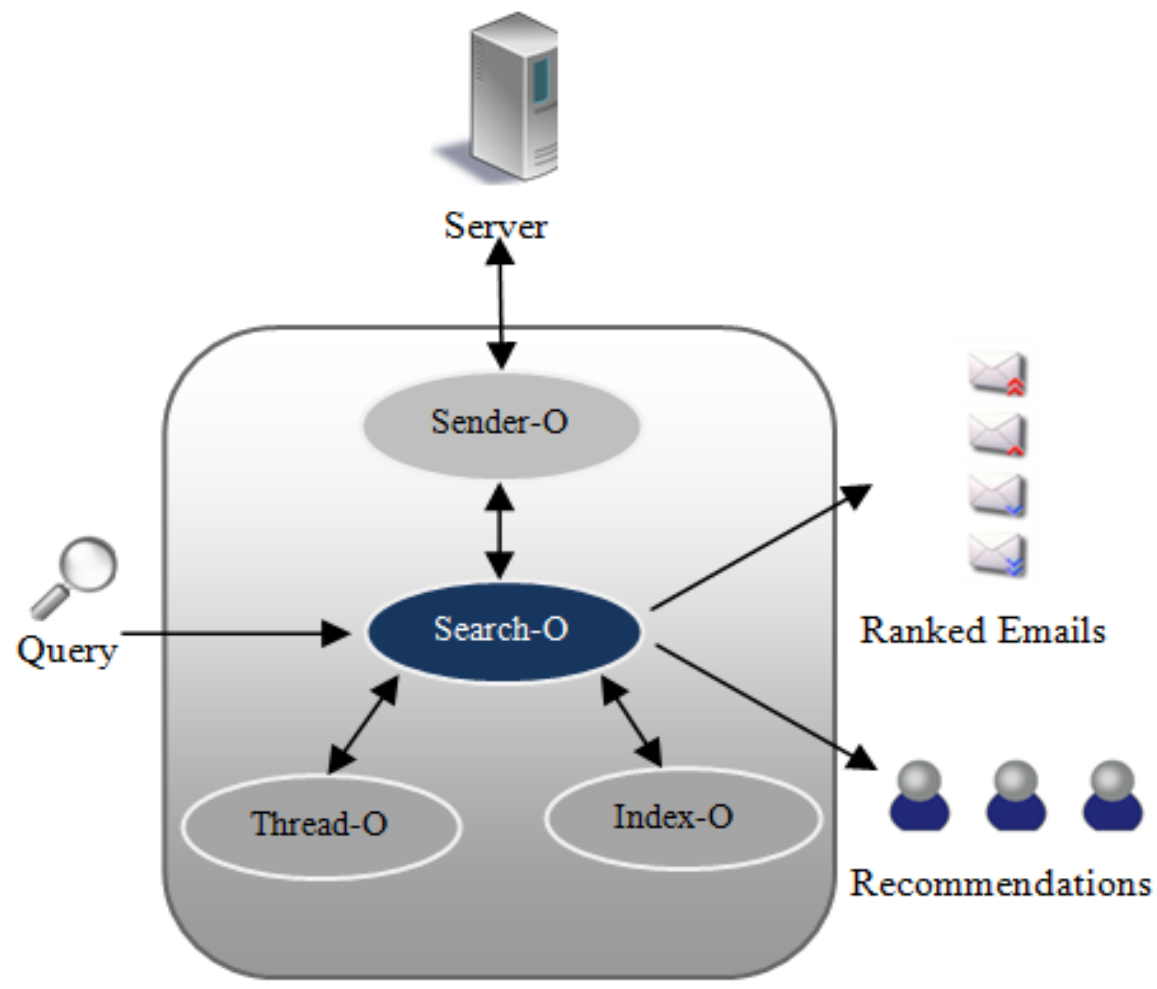

Figure 8. The Proposed Email Client Architecture

## 6. EXPERIMENTAL EVALUATION

The IR system evaluation depends on two main factors namely, the test collection and the evaluation measure(s). This section first demonstrates the test collection setup and states the used evaluation measure. Finally, it describes and analyzes the experimental results.

### 6.1. Test Collection Setup

The test collection used for IR system evaluation mainly consists of three elements: (1) a document collection, (2) queries or the set of information needs, (3) the relevance judgment. Up to our knowledge, no standard test collection is currently publicly available for Email Retrieval Ranking research. Therefore, we set up our own test collection manually. In the next subsections, the related details are described.

#### 6.1.1. Document Collection

In our experiments, Enron email corpus was used [7]. At first, Enron email dataset was made public by the Federal Energy Regulatory Commission (FERC), and then later the Cognitive Assistant that Learns and Organizes (CALO) project collected and prepared a new dataset version, which was made available for research. This version contains about 517, 431 emails (~1.32 GB) owned by about 151 users of Enron Corporation; most of these emails are senior management staff. The emails are spread over 3500 folders [13].

We filtered the mentioned dataset by selecting only two principal folders “sent_items” and “inbox” from each user email directory having these two folders. All other folders were created either by the user or by the system for email categorization task. Our filtered dataset version contains around 80,110 emails (~221 MB) for 137 users distributed among 410 folders.

Then the 137 users above were categorized into four classes, according to the total number of emails contained in their principal folders. The four classes are large, medium, small, and micro, in which the total number of emails is > 1500, between 1000 and 1500, between 500 and 1000, and < 500 respectively. Figure 9 shows that the majority of users belong to micro and small classes, 82% of users, and the minority ones belong to the other classes, only 18% of users.

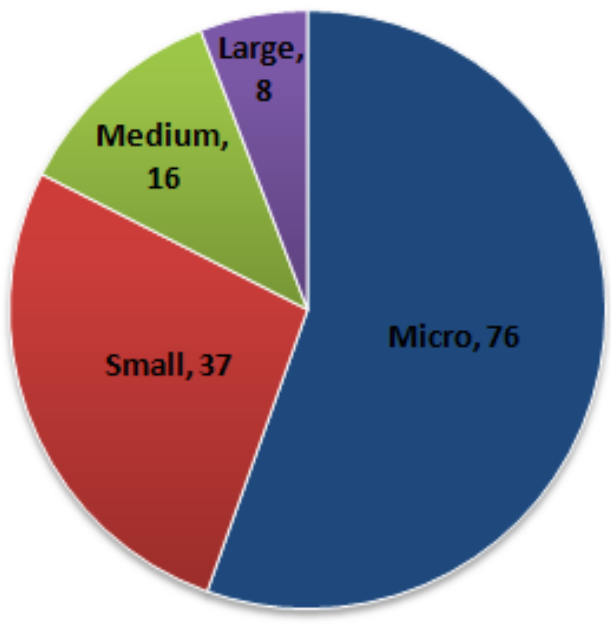

Figure 9. The Number of Users in Each Class

Figure 10 depicts some essential information about the large class users; on which we focused. For each user, this figure lists the Enron email directory name and the total number of its emails. Among these users, we randomly selected two users to present our document collection namely, *Steffes-J* (Vice President of Government Affairs) and *Sager-E* (Employee).

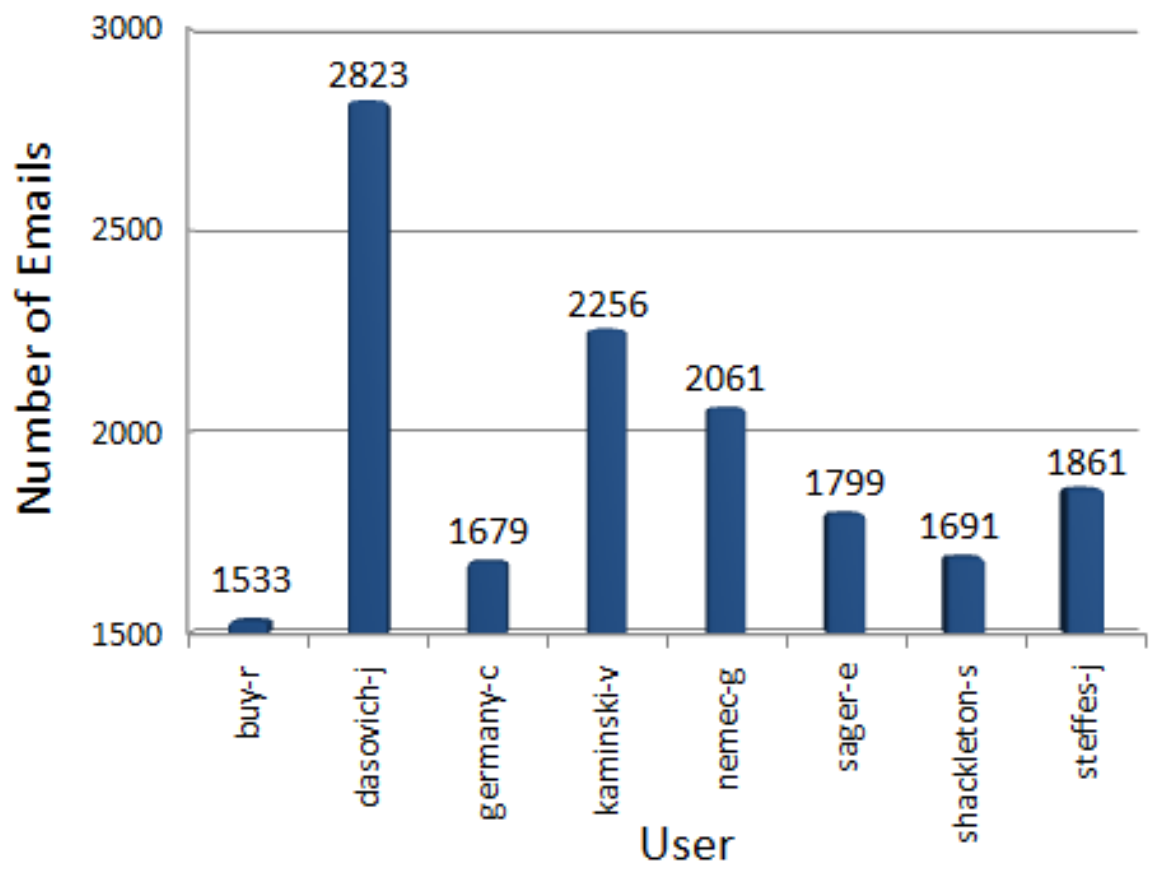

Figure 10. The Number of Emails for Large Class Users

### 6.1.2. Queries

We adapted a set of 300 Enron queries gathered from different resources, such as a set of categories developed by UC Berkeley Enron Email Analysis Project [14], 2001 annotated Enron Email Dataset topics [15], and Trampoline Enron Explorer system themes. In this paper, a subset of this query set was used. 35 queries were randomly selected from our query set; Table 1 lists these queries according to their length, in terms of the number of words, i.e. one-word queries, two-word queries, and three-word queries. When these queries were applied on our email dataset, the largest retrieved set contained 648 emails, and the smallest one contained 7 emails. On the average, 50 emails were retrieved per query.

Table 1. Set of the Queries Selected for the Experiments

| ***One Word*** | ***Two Words*** | ***Three Words*** |
|---|---|---|
| Bankruptcy | Master Netting | California Energy Crisis |
| Tennis | Contract Review | Happy New Year |
| Collateral | Houston Meeting | North American Energy |
| FERC | Conference Call | |
| Spreadsheet | Bank Account | |
| Settlement | Budget Meeting | |
| Letters | California Power | |
| NYISO | Annual Fees | |
| Attachment | Call Me | |
| Bill | Financial Assurance | |
| Presentation | Dynegy Company | |
| Requirement | Energy Marketing | |
| Taxes | Expense Report | |
| Complaints | Federal Government | |
| | Government Affairs | |
| | Market Power | |
| | Meeting Schedule | |
| | Gas Transportation | |
| | Project Progress | |

### 6.1.3. Relevance Judgment

Relevance judgment or assessment is the most effective factor in the test collection for IR system evaluation. In any relevance judgment, each corpus document is assessed according to its relevance to the submitted query. This can be done in two ways, either as binary or graded relevance assessments. While the former method judges each document as being either 'relevant' or 'irrelevant' to the query, the latter one weighs it according to its relevance grade to the query. The amount of information provided by each relevant email is not similar; for example, one may discuss the query theme in more depth in one email, and he may cover more or other aspects of it in the other. Therefore, we used graded relevance assessment, on the retrieved emails only, to obtain more accurate evaluation for the proposed ranking approach. In our assessment approach, the pooling method was applied to acquire the top 100 emails for each query from each Email Retrieval system, which was used in our experiments, including the proposed approach. Then, the pooled emails were manually assessed by three judges. For each query-email pair, the relevance degree level that is assigned to the email reflects its relevance to the query, based on the information contained in the email. The assessment was done on 4-point scale levels 0, 1, 2, and 3 corresponding to irrelevant, marginally relevant, fairly relevant, and highly relevant respectively. These cases are detailed as follows:

- *Irrelevant*:
  - When the email does not contain any information about the query theme.
  - When the query statement is mentioned in the email content but it refers to a different theme; i.e. they both are semantically different. For example, the email which talks about 'Bill', a person name, is not related to the query 'Bill' referring to the statement of money; although, the word 'Bill' is mentioned frequently in such an email.
- *Marginally relevant*:
  - When the email just contains the query statement, in the main body (unquoted text), but it does not contain more information about the query theme.

- When the email contains information about the query theme in few quotations.

- *Fairly relevant*:
    - When the email contains more information related to the query theme rather than the query statement but this information is moderate; i.e. 40 to 80% of the information gained from its main body (unquoted text) is related to the query theme.

- *Highly relevant*:
    - When the email theme exhaustively discusses the query theme, i.e. more than 80% of the information gained from its main body (unquoted text) is related to the query theme.
    - When the query statement is contained in the email subject.

## 6.2. Evaluation Criteria

Precision-Recall curve, Mean Average Precision (*MAP*), and Precision at the $K^{th}$ ranked position (*P@K*) are all measures for evaluating the effectiveness of the ranked retrieval results [1]. These measures depend in their calculation mainly on Precision and Recall measures [1], which in turn depend on the binary relevance assessment of the documents. Another popular measure is Normalized Discounted Cumulative Gain (*NDCG*) [16]. The measure is based on two facts. First, highly relevant documents are more useful to the user than marginally relevant ones. Second, the user usually checks the fewest top documents of the retrieved list. Therefore, *NDCG* was used to evaluate the proposed ranking approach because it uses graded relevance. For the submitted query, *NDCG* is calculated at the $K^{th}$ ranked position of the retrieved list (*NDCG@K*) (Equation. (9)).

$$DCG@K = \sum_{i=1}^{K} \frac{2^{rel(i)} - 1}{\log_2(1+i)} \quad (8)$$

$$NDCG@K = \frac{DCG@K}{DCG_{ideal}@K} \quad (9)$$

Where *rel(i)* is the relevance grade associated to the document at position *i* of the retrieved list for the given query. *DCG* values (Equation (8)) are divided by the *DCG* values of the ideal rank ($DCG_{ideal}$) to obtain *NDCG*. The ideal rank is acquired by sorting the retrieved list in descending order according to the relevance grade of its documents, such that the high relevant documents locate on top of the list. Therefore, all *NDCG@K* values always fall between 0 and 1 regardless of the retrieved list size. *NDCG* was computed for the top 100 documents (*NDCG@1-100*) in our experiments. In the following section, our analysis focuses mainly on the top 10 documents (*NDCG@1-10*), since this is usually the number of documents that the user affords to examine from the retrieved list. Moreover, as a sample results for *NDCG@1-10*0 experiments, Figure 12 shows evidence that our approach is highly effective for the top 100 documents.

## 6.3. Experimental Results

Up to our knowledge, there is no baseline system that employs email IR ranking approaches to compare our approach with. Therefore, to evaluate the performance of our proposed Email Ranking Approach (ERA), its effectiveness is compared with that of current popular commercial email systems and some other email ranking approaches. The evaluation is done using NDCG measure and our test collection; as being mentioned in the previous sections.

Our experimental study is performed through three experiments. The first experiment (Experiment1) studies the performance of our proposed ERA compared to that of the current commercial email systems. The second experiment (Experiment2) studies the performance of our proposed ERA compared to the current email ranking approaches and the approaches presented in previous researches. Finally, the third experiment (Experiment3) studies the improvements that can be added when applying the proposed networking architecture to the proposed ERA.

### 6.3.1. Retrieval System Quality

A java open source email client named *Columba* is used to implement our proposed ERA and some other ranking approaches that we investigated in our experiments.

Our implemented email retrieval system applies query expansion methods using *Porter Stemmer* and *Edit distance* (Levenshtein distance) algorithms. Stemming is used to retrieve the emails containing terms with stems similar to the query terms stem, which may be relevant to the submitted query. Since emails usually contain misspelled words; therefore, Edit distance algorithm is used to retrieve the emails containing misspelled query terms. Table 2 lists sample query terms and their corresponding expanded terms obtained from edit distance and Porter stemmer algorithms. This expansion increases the recall but may harm the precision. However, our proposed ranking approach tackles this by presenting most relevant emails to the user on top of the list, and it lists irrelevant ones at the end of list.

Table 2. Query Expansion for Some Query Terms

| *Query Term* | *Edit Distance* | *Stemming* |
|---|---|---|
| Attachment | attachements | attach, attached, attaches, attaching, attachments |
| Account | accoint | accountability, accountable, accountants, accounted, accounting, accounts |
| Budget | busget | budgets, budge |
| Company | compny | companies |
| Government | governement | govern, governance, governments, governmental |
| Meeting | meeing | meet, meetings, meets |
| Requirement | requeriments, requriements | require, required, requirements, requires, requiring |

We focus in this paper on ranking the retrieved list, but we conduct an experiment to ensure that the quality of the retrieved list by our system is nearly similar to the lists retrieved by the compared systems. Therefore, *Recall* (Equation (10)), *Precision* (Equation (11)), and *F-Measure* (Equation (12)) [1] are calculated for Gmail, Windows Mail, and our system. To calculate these measures, using the graded relevance judges, which we use, the 4 levels are categorized into two levels only, namely relevant and irrelevant. In which, all positive relevant levels, i.e. 1, 2, and 3, are considered relevant, while level 0 is considered irrelevant. Moreover, all the documents that are not judged are considered irrelevant.

$$\mathrm{Re}call = \frac{\#\, relevant\ emails\ retrieved}{\#\, relevant\ emails} \tag{10}$$

$$\mathrm{Pr}ecision = \frac{\#\, relevant\ emails\ retrieved}{\#\, retrieved\ emails} \tag{11}$$

$$F - Measure = 2 * \frac{Recall * Precision}{Recall + Precision} \tag{12}$$

Table 3 demonstrates the average Recall, Precision, and F-Measure values of Gmail and Windows Mail systems compared with our system (ERA) on 35 queries. From the Table, it is noticed that ERA achieves the best recall, but its precision is lowest. The reason behind this is the use of query expansion in our system. However, ERA has the best F-Measure score.

Table 3. Average Recall, Precision, and F-Measure values

| *Email System* | *Recall* | *Precision* | *F-Measure* |
|---|---|---|---|
| Gmail | 0.769 | **0.733** | 0.718 |
| Windows Mail | 0.832 | 0.708 | 0.736 |
| ERA | **0.913** | 0.707 | **0.771** |

### 6.3.2. Experiment1: "Email Systems"

In this experiment, the effectiveness of the two well-known email systems, namely *Gmail* and *Windows Mail* are evaluated.

- ***Gmail***: is thread (conversation) oriented; it groups related emails by subject into a thread, where emails are stacked in the thread. When an email is relevant to the submitted query, it retrieves the whole thread in the retrieved list, and the system sorts the retrieved threads in descending chronological order.
- ***Windows Mail***: like all traditional email client systems, its default retrieval ranking method is the reverse chorological order in which the newest emails appear on the top of the retrieved list.

Figure 11 reports the experiment results; the experiments were conducted on the two users email data namely, *Sager-E* and *Steffes-J*. Figures 11(a) and 11(b) report NDCG average values of 25 queries and 11 queries respectively. It is clear from the figures that ERA outperforms both Gmail and Windows Mail systems at all ranks and in both user email data. In Figure 11(a), ERA improves NDCG values of Gmail at ranks 1, 2, and 3 by 31.4%, 29.1%, and 23.7% respectively; and, it improves Windows Mail NDCG values by 31.5%, 29.3%, and 28.4% respectively. Similar improvements are observed in Figure 11(b); while ERA improves NDCG of Gmail at ranks 1, 2, and 3, by 42.9%, 46.9%, and 50.2% respectively, NDCG values of Windows Mail are boosted by 45.5%, 48.5%, and 52.3% respectively.

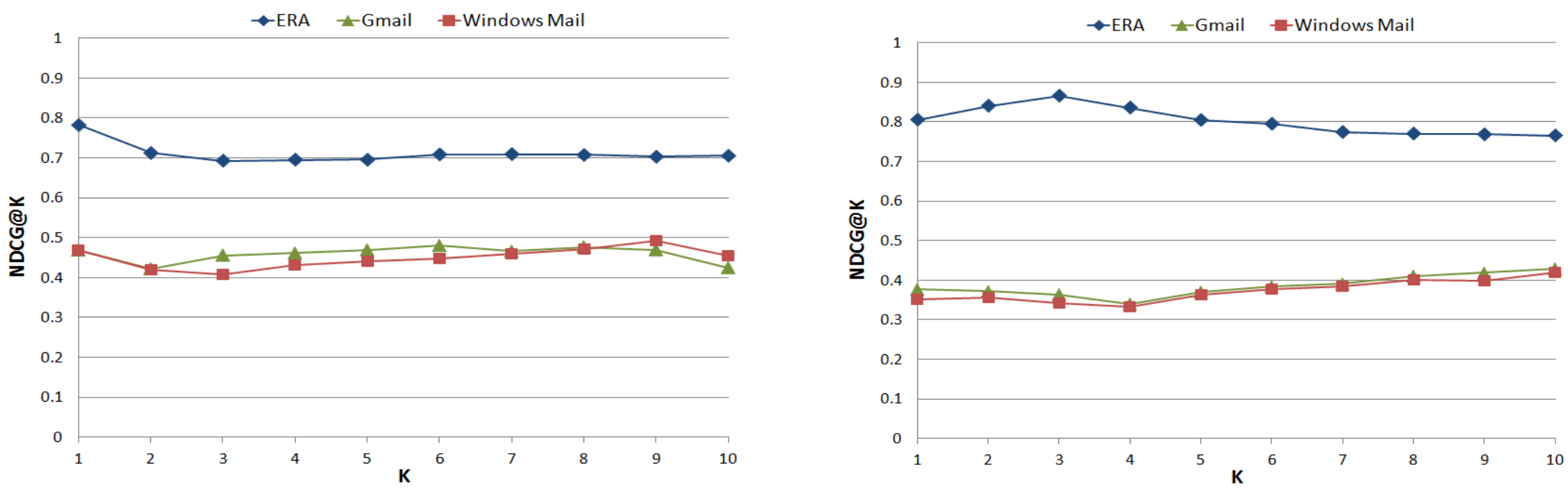

(a) Steffes-J Email Data (b) Sager-E Email Data

Figure 11. Average NDCG@1-10 of ERA and Email Systems

Figure 12 shows NDCG average values of the proposed ERA and the two systems at some selected ranks from rank 1 to 100; these ranks are 20, 40, 60, 80, and 100. All these values report that ERA is superior to both Gmail and Window Mail in all the mentioned ranks.

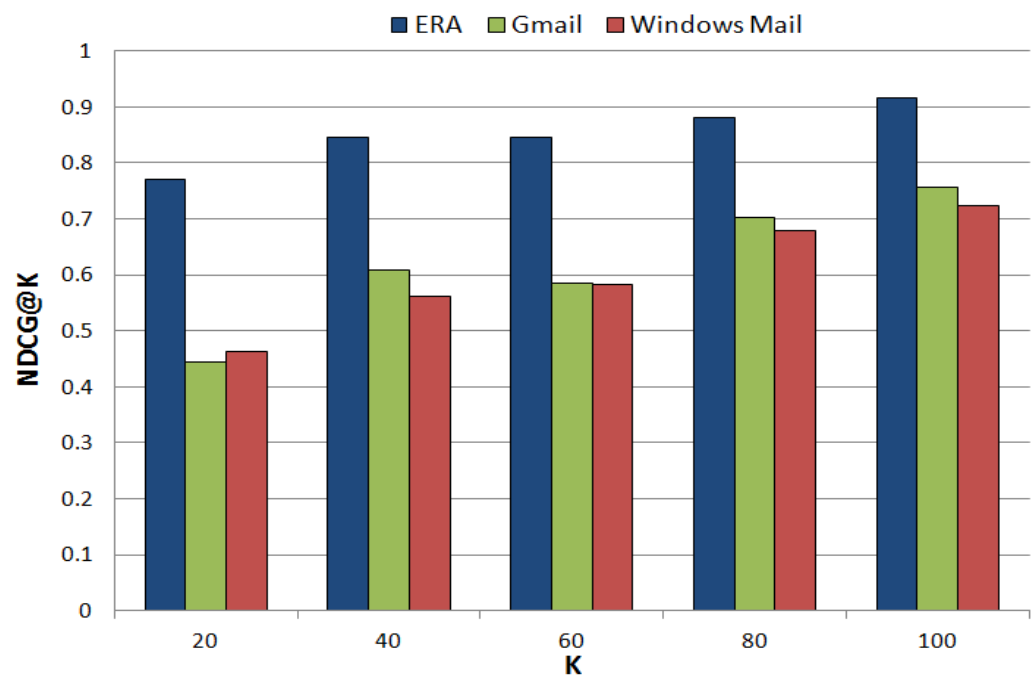

Figure 12. Average NDCG@K of ERA and Email Systems

Since the user always hopes to navigate first what he requires, we must give attention to the improvements that our proposed ERA achieves at first rank, i.e. NDCG@1. Figure 13 reports NDCG@1 values for the 11 queries of the proposed ERA and the other two systems. ERA outperforms the other systems in the 11 queries. Moreover, ERA reaches the ideal rank, i.e. NDCG@1=1, in 8 queries from 11, while Gmail and Windows Mail reach only the ideal rank in 2 queries from 11. The above experiment results conclude that our proposed ERA surpasses Gmail and Windows Mail systems.

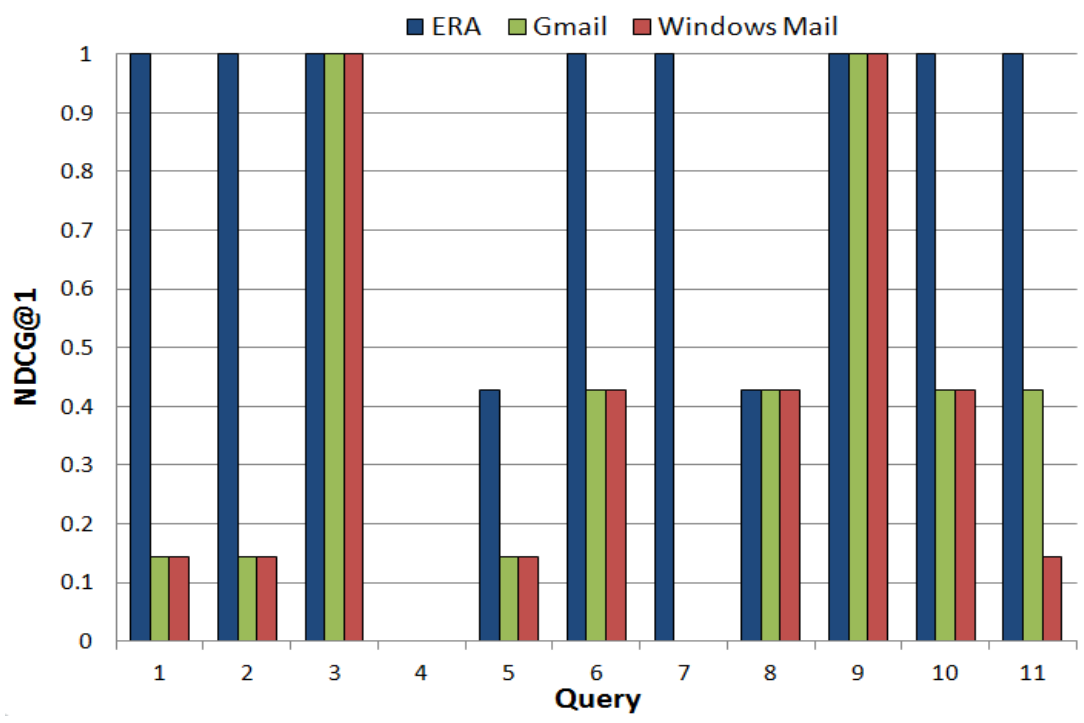

Figure 13. NDCG@1 for 11 Queries of ERA and Email Systems

### 6.3.3. Experiment2: "Ranking Approaches"

In this experiment, the effectiveness of email ranking approaches that depend on predefined user sorting criteria are evaluated. We investigate the following methods:

- ***Subject***: the retrieved emails are sorted alphabetically in descending order according to the email subject.
- ***Sender***: the retrieved emails are sorted alphabetically in descending order according to the email sender; either email address or name.
- ***Clues***: the retrieved emails are ranked based on the clues, which are submitted with the user query. In this experiment, the approach presented by Perkiö et.al [7] was applied. Therefore, the emails that have more words that frequently co-occur with the clue words are placed on top of the retrieved list.

This experiment was conducted on Sager-E data using Experiment1 queries. Figure 14 reports the results of this experiment. We deduce that subject based ranking outperforms both sender based and clues based approaches at all ranks. Fortunately, our proposed ERA is superior to Subject method in all ranks. At ranks 1, 2, and 3, ERA improves NDCG Subject ranking values by 35.1%, 40.1%, and 43.8% respectively.

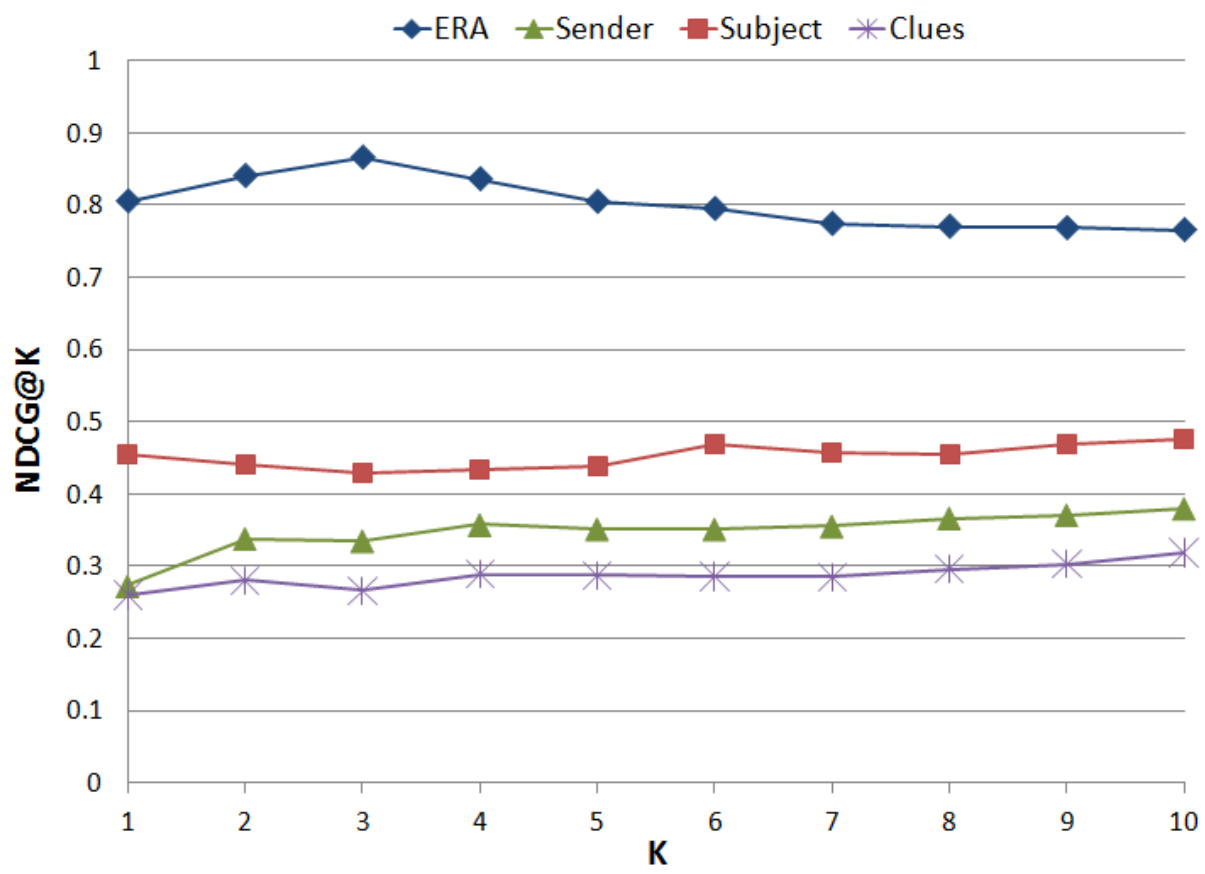

Figure 14. Average NDCG@1-10 of ERA and Ranking Approaches

### 6.3.4. Experiment3: "The Proposed Architecture"

In this experiment, we study the effect of applying the proposed networking architecture on our proposed ERA. The proposed architecture was applied on a small scale Enron users email data, in which Sager-E data presents the local e-mailbox side. To gather the interests of all network users, a simple server is proposed to handle all communications and transactions of the senders through the network; we let each user submits, to the server, the subset of his sent emails that are permitted for public, to simply handle some network security issues.

Figure 15 demonstrates the average NDCG values of 11 queries for Local ERA and Global ERA. While the former is typically the proposed approach, the latter is the network ERA. The figure shows that Global ERA slightly improves Local ERA at some ranks. For example, at rank 2 it improves NDCG by 2% and almost by 1% at ranks 7, 8, 9, and10. At other ranks, Local ERA gains better NDCG values, like NDCG@4 and NDCG@6.

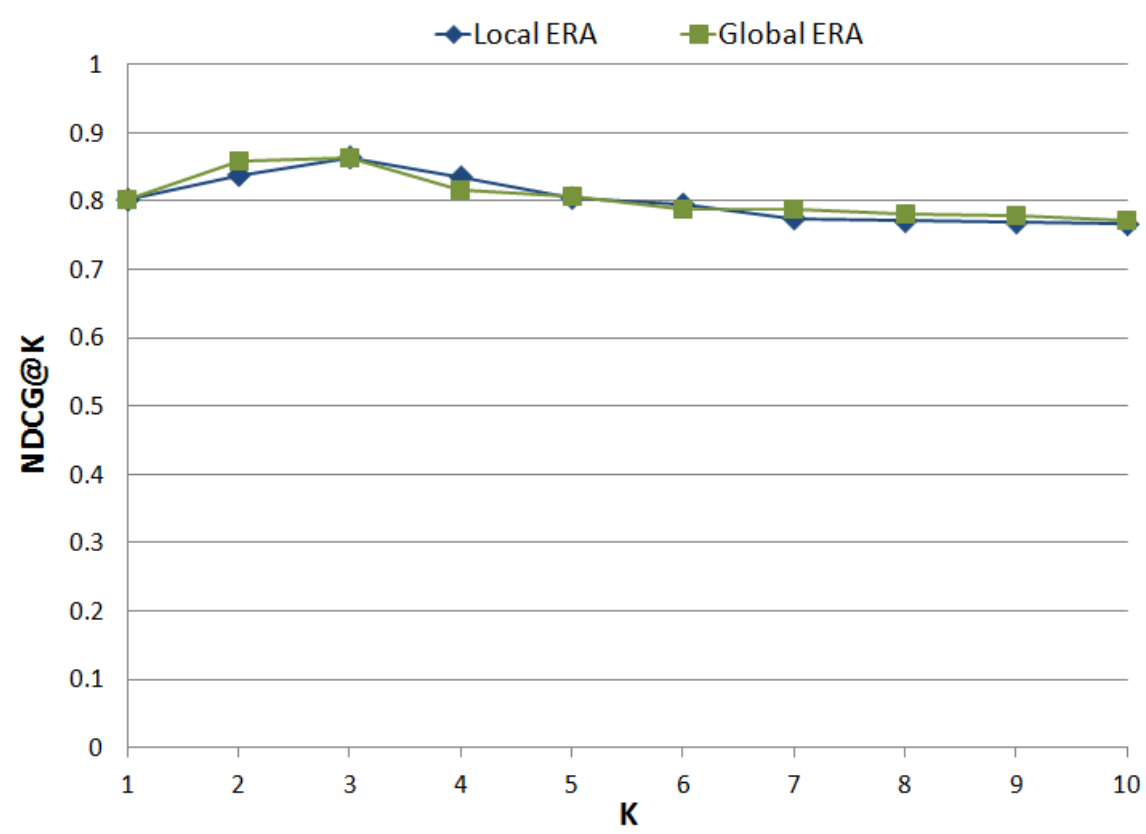

Figure 15. Average NDCG@1-10 of Local and Global ERA

The reasonable analysis of the above results is that, from the e-mailbox owner view, some senders may be considered specialists in some themes if they mostly communicate with him in such themes. However, these senders may not actually be real theme specialists over the network. Therefore, when Local ERA is applied, the emails sent by these senders get higher scores than Global ERA ones; these are the cases at ranks 4 and 6 (Figure 15). Moreover, the e-mailbox owner may have fewer communications with a sender about some themes; however, this sender may be the themes specialist over the network. If the emails of this sender are highly relevant to the query, then they get low scores when Local ERA is applied, and they get high scores if Global ERA is applied. This is the case at rank 2 (Figure 15).

From the above results, one can conclude that both approaches may help the user to find the theme specialist to contact him when needed. Therefore, there is no absolute best approach. However, there always exists the most adequate approach in which the user may achieve his desires, given that it is permissible by the related network security protocol. According to this criterion, the network server may offer the names of the theme specialists to all network users. The user may then utilize this helpful knowledge to adjust the weights of his contact list senders, if any, for more accurate email ranking. He may even consider it as just good recommendations to carry out when needed. For example, Elizabeth Sager, Sager-E data owner, may benefit from our proposed system recommendation lists for her submitted queries (Table 4); the lists concerned of the network users who frequently communicate with others regarding the query search theme.

Table 4. Recommended Users for Some Search Queries

| ***Query*** | ***Recommended Users*** |
|---|---|
| Master Netting | Alan Comnes, Christian Yoder, James Steffes |
| FERC | Sarah Novosel, Alan Comnes, Robert Frank |
| Houston Meeting | James Steffes, Jennifer Thome, Ginger Dernehl |
| Bankruptcy | Vicki Sharp, Jeff Dasovich |
| Collateral | Juan Padron, Harry Kingerski |

To conclude our experiments, we measured the average time of selective 31 queries among all queries that are prepared for the experiments. The purpose of this study is to ensure that our approach user response-time is reasonable. The average time is 750 msec for average 55 retrieved emails, using Intel(R) Core(TM)2 Duo CPU @ 2.53 GHz processor. This time is considered reasonable for the following reasons:

1) This research is interested in the Email Retrieval Ranking effectiveness. Therefore, we did not make any implementation optimizations to improve the research efficiency.
2) The communications among network objects themselves and the server increase the response-time, and it is needed to be optimized.

In spite of the above response-time optimizations, we do believe that the user may afford to have high retrieval response-time with accurate email ranking close to his thinking and he can not afford to have low retrieval response-time with chaos ranking far from his expectation.

## 7. Conclusions And Future Work

In this paper, we presented a new Email Retrieval Ranking approach. Unfortunately, the current Email Retrieval Ranking approaches do not satisfy the user needs, since their rankings are based on predefined criteria, such as some email fields, or some search clues that are determined by the user, which are query independent features. Our proposed approach ranks the retrieved emails, based on some basic real-life user heuristics. It utilizes simple scoring function based on subject, content, and sender email fields, in which, the score of each retrieved email is the sum of the content and subject fields similarity scores to the query weighted by the email sender score based on his specialty in the query theme. In addition, it uses some network information regarding user interests to promote the ranking of the retrieved emails and to offer some contacts of specialists in the submitted query theme, as recommendations to the user.

We carried out some empirical evaluations on two users email data, from Enron email dataset, whose emails are considerably large. The results showed that our proposed Email Retrieval Ranking approach outperforms the current literature ranking approaches. Hence, we proved that our approach reflects the email user thinking in ranking the retrieved emails.

In future, we plan to investigate other email fields, coupled with suitable people heuristics, to cope with our approach. We also will study how to get benefit from the user profile to enhance the ranking scores or recommendations.